\documentstyle[12pt]{article} 

\def\s{\scriptstyle}
\def\ss{\scriptscriptstyle}

\def\ds{\displaystyle}
\newcommand{\be}{\begin{eqnarray}}
\newcommand{\ee}{\end{eqnarray}}
\newcommand{\nn}{\nonumber}

\def\slashxi{{\xi}\!\!\!/} 
\def\nx{n(\xi,t)}

\def\o{\over}
\def\C{{\s C}}

\def\pmic{P(\C_i,t)}

\def\pmicl{P(\C_i,t+1)}

\def\pmicf{P'(\C_i,t)}
\def\pmicjf{P'(\C_j,t)}
\def\pmicc{P_c(\C_i,t)}
\def\pmicjc{P_c(\C_j,t)}

\def\pmicLf{P'(\C_i^{\ss L},t)} 
\def\pmicRf{P'(\C_i^{\ss R},t)}

\def\pmicLLf{P'(\C_i^{\ss LL},t)} 
\def\pmicLRf{P'(\C_i^{\ss LR},t)}

\def\pmicL{P(\C_i^{\ss L},t)} 
\def\pmicR{P(\C_i^{\ss R},t)}
\def\px{P(\xi,t)}
\def\pxl{P(\xi,t+1)}

\def\pxf{P'(\xi,t)}

\def\pxc{P_c(\xi,t)}
\def\pnxc{P_c(\slashxi_i,t)}
\def\pnx{P(\slashxi_i,t)}
\def\pxLf{P'(\xi_{\ss L},t)}
\def\pxRf{P'(\xi_{\ss R},t)}
\def\pxLft{P'(\xi_{\ss L},t')}
\def\pxRft{P'(\xi_{\ss R},t')}

\def\fav{{\bar f}(t)}
\def\fxi{{\bar f}(\xi,t)}
\def\frat{{\fxi\over \fav}}
\def\fx{{f_{\xi}}}
\def\fxL{{f_{\xi_L}}}
\def\fxR{{f_{\xi_R}}}
\def\fmic{f(\C_i,t)}

\def\dfx{\delta\fx}
\def\dfxL{\delta\fxL}
\def\dfxR{\delta\fxR}

\def\muti{{\cal P}({\ss C_i\ra C_i})}

\def\mutji{{\cal P}({\ss C_j\ra C_i})}
\def\mutis{{\cal P}({\s \xi\ra\xi})}
\def\mutijs{{\cal P}({\s \slashxi_i\ra\xi})}
\def\crossij{{\cal C}_{\ss C_iC_j}^{(1)}(k)}
\def\crossjl{{\cal C}_{\ss C_jC_l}^{(2)}(k)}
\def\hamij{d^H(i,j)}
\def\hamijL{d^H_L(i,j)}
\def\hamijR{d^H_R(i,j)}
\def\hamilL{d^H_L(i,l)}
\def\hamilR{d^H_R(i,l)}
\def\ra{\rightarrow} \def\lra{\longrightarrow}
 \def\i{\infty}
\def\ef{f_{\ss\rm eff}(\xi,t)}
\def\efff{f'_{\ss\rm eff}(\xi,t)}
 \def\seff{s_{\ss\rm eff}}

\begin{document}
 
\title{Schemata Evolution and Building Blocks} 
 
\author{ {\bf Chris Stephens \thanks{e-mail: stephens@nuclecu.unam.mx}
\ {\rm and} Henri Waelbroeck 
\thanks{e-mail: hwael@nuclecu.unam.mx}} \\  
Instituto de Ciencias Nucleares, UNAM\\   
Circuito Exterior, A.Postal 70-543 \\ 
M\'exico D.F. 04510 \\} 
\date{12th October 1997} 
\maketitle 
  
\begin{abstract} 
In the light of a recently derived evolution equation for genetic algorithms
we consider the schema theorem and the building block hypothesis.
We derive a schema theorem based on the concept of 
{\it effective fitness} showing that schemata
of higher than average effective fitness receive an exponentially increasing 
number of trials over time. The equation makes manifest the content of
the building block hypothesis showing how fit schemata are constructed
from fit sub-schemata. However, we show that generically there is no 
preference for short, low-order schemata. In the case where
schema reconstruction is favored over schema destruction
large schemata tend to be favored. As a corollary of the evolution equation
we prove Geiringer's theorem.
\end{abstract} 
\vskip 0.4truein

\section{Introduction} 

One of the most commonly asked questions about genetic algorithms (GAs) is:
under what circumstances do GAs work well? Obviously an answer to this 
question would help immeasurably in knowing to which problems one can apply
a GA and expect a high level of performance. However, to answer this
question one has to answer a more fundamental question: how do GAs work? 
For example, in a typical optimization problem how does the GA arrive at
a good solution? It is clear that in very complex problems this is not
achieved via a random search in the state
space. The search is structured. However, the question remains as to what 
is the nature of this structure. To put this question another way, if we
think of individual string bits as ``degrees of freedom'', the GA does not
exhaustively search through the different combinations of individual
bits, i.e. a search in the entire state space. Rather it 
searches through a restricted space spanned by different combinations of
``effective degrees of freedom'' (EDOF), which are combinations of the more
fundamental ``microscopic'' bit degrees of freedom. 

What exact form these EDOF take depends of course on the 
particular landscape under consideration. Hence, one might despair as to 
whether it was possible to say anything that applied to more than a
specific case. However, it is not 
meaningless to try to understand if they exhibit generic properties, 
independent of the landscape, or at least properties that are common
to a large class of possible landscapes. The building block 
hypothesis and the schema theorem 
\cite{holland}, \cite{goldberg} 
attempt to identify such
generic features and as such have played an important role in GA 
theory, if one accepts that one of the principal goals of a theory 
is to provide a framework within which one can gain
a qualitative understanding of the behavior of a system. The basic 
gist of the building block hypothesis is that {\it short, low-order}, 
highly fit schemata play a preeminent role in the evolution of a GA; i.e.
that the relevant EDOF for a GA are 
short, low-order, highly fit schemata. The schema theorem tries to lend
a more quantitative aspect to the hypothesis by showing that 
such schemata are indeed favored. This fact is deduced via an analysis of the 
destructive effects of crossover. However, as is well known, the
schema theorem is an inequality and is such because it does not say
anything precise about schema reconstruction. To understand better
the interplay between schema destruction, schema reconstruction
and schema length one requires an evolution equation that is exact,
and where schemata are the fundamental objects considered.

Various exact evolution equations have been derived previously: 
\cite{gold87} 
wrote down exact equations for two-bit problems. Later
these equations were extended to three and four-bit problems 
cite{whitley}.
These equations allowed for an explicit analysis of string gains and losses.
\cite{whitcrab} 
also presented an algorithm for 
generating evolution equations for larger problems that was equivalent 
to an earlier equation of Bridges and Goldberg 
\cite{goldbridge}. 
Although exact these equations are extremely unwieldy and it is difficult to
infer general conclusions from their analysis. Another related approach is that
of Vose and collaborators 
\cite{vose}, \cite{vosee}, \cite{vose3} 
that treats GA evolution as a Markov chain. One of the chief drawbacks of all the above, 
with respect to an analysis of the schema theorem and the block hypothesis, is that the former
are evolution equations for strings whereas the latter refer to schemata. 
Evidently an evolution equation that is amenable to interpretation and 
analysis that treats schemata as fundamental objects would be  
preferable. Such an equation has been derived recently \footnote{After the completion
of this work we became aware of the related work of Altenberg \cite{alt}.}
\cite{stewael}, \cite{stewaell} 
for the case of proportional selection and 1-point crossover. 
The chief aim of this paper is to analyze the schema theorem and the 
building block hypothesis in the light of this equation.

Crucially, we will be able to quantify the effect of schema reconstruction
relative to that of schema destruction. Traditionally, crossover as
a source of schema disruption has been emphasized 
\cite{disrup}, \cite{disrupp}. 
This idea is at the heart of the schema theorem and the building 
block hypothesis. There has been some work towards a more positive 
point of view of crossover vis a vis reconstruction 
\cite{recon}, \cite{reconn} 
but mainly in the light of the exploratory nature of crossover. 
Here, we will see exactly under what conditions schema reconstruction 
dominates destruction.

In analyzing the consequences of the evolution 
equation we will especially emphasize two ideas: effective
fitness and EDOF. With respect to the former we will show
that: if one thinks intuitively of fitness as representing the ability of a 
schema to propagate then effective fitness is a more relevant concept
than the conventional idea of fitness. We will formulate
a schema theorem in terms of the effective fitness showing that 
schemata with high effective fitness receive an exponentially 
increasing number of trials as a function of time.
The second key idea, already mentioned, is that of EDOF. 
Generically one can think of a schema as an EDOF. However, 
schemata offer for every string a decomposition into $2^N$ 
different elements of a space with $3^N$ members.
Not every decomposition will be useful. In fact, typically, only
a small subset. So what do we mean by useful? 
EDOF, if they are to have any utility whatsoever, should not 
be very strongly coupled. This is a notion that is intimately 
associated with how
epistasis is distributed in the problem. This type of thinking
is common to many fields and generally is associated with 
the idea of finding a basis for a highly non-linear problem 
wherein it decomposes into a set of fairly independent 
sub-problems. An important feature of complex systems is that
the EDOF are ``scale'' dependent. This scale dependence very
often takes the form of a time dependence wherein the EDOF are
different at different stages of evolution. This
complicates life greatly in that if we find a useful decomposition
of a problem at time $t$ we have no guarantee that it will remain
a useful decomposition indefinitely into the future.

\section{Coarse Graining and Schemata}

As is well known to any scientist or engineer a good model of a system is
one that captures the relevant features and ignores irrelevant details. 
The deemphasis of irrelevant details we can think of as a ``coarse graining''.
Of course, a great difficulty is that often what is relevant versus 
irrelevant depends on what one wants to say about the system, i.e. what 
level of description one requires. It also, more often than not, depends on time.
One of the most obvious examples of this is evolution: the primitive 
constituents of life, amino acids, DNA, RNA etc, which represent the 
``microscopic'' degrees of freedom, over time combined to form progressively
more and more complicated EDOF such as cells, sponges, people etc. This 
evolution in time is intimately linked to an evolution in ``scale'' and
a corresponding evolution in complexity. 

In a GA specifying all the bits of a string gives us the most fine grained,
microscopic description possible. For strings of size $N$ and a population
of size $n$ there are $Nn$ degrees of freedom and, for a binary alphabet, 
$O(n2^N)$ possible population states. 
Consider the different classes of fitness maps that may be defined: 
first, $f_G:G\lra R^+$, where $G$ denotes 
the space of genotypes (string states) and $f_{\ss G}$ is the fitness function
that assigns a number to a given genotype; second, $f_{\ss Q}:Q\lra R^+$, where 
$Q$ is the space of phenotypes. These mappings 
may be explicitly time dependent. In fact, this will normally be the case 
when the ``environment'' is time dependent. 
They may also be injective or not, although the map $f_{\ss Q}$ will
usually be injective. If $f_{\ss G}$ is many-to-one then there 
exist ``synonymous'' genotypes, i.e.
the mapping is degenerate. If we assume there exists a map 
$\phi:G \lra Q$ between genotype and phenotype then we 
have $f_{\ss G}=f_{\ss Q}\circ\phi$,
i.e. the composite map induces a fitness function on $G$.
A schema, $\xi$, consists of $N_2\leq N$
defined bits. The defining length of the schema, $l$, is the distance between 
its two extremal defining bits. The space of all schemata, $S$, 
may be partitioned according to 
schema order; i.e. $S=\sum_{N_2}S_{N_2}$, where $S_{N_2}$ is the space of
schemata of order $N_2$. The mapping $g:G\lra S$ between strings and schemata
is many-to-one. The degree of degeneracy of the map, $g_{N_2}:G\lra S_{N_2}$, is
$2^{N-N_2}$. Except for the trivial case of a $0$-schema, maximal degeneracy 
occurs when $N_2=1$ where half of $S$ is mapped onto one schema.
The fitness of a schema is the map $f_S:S\lra R^+$, which is related to
$f_{\ss G}$ via the composite map $f_S\circ g=f_G$. Explicitly,
\begin{eqnarray}
\fxi= {\sum\limits_{\C_i\supset\xi}\fmic n(\C_i,t)  \o 
\sum\limits_{\C_i\supset\xi}n(\C_i,t)}\label{scfit}
\end{eqnarray}
where $f(\C_i,t)$ is the fitness of string $\C_i$ at time $t$, $n(\C_i,t)$ is
the expected number of strings of type $\C_i$ at time $t$ and the sums
are over all strings in the population that contain $\xi$.

As mentioned, the total number of schemata for a binary alphabet 
is $3^N$. Why go to an even bigger space than the state space itself? 
One answer to this 
question is related to the idea of coarse graining. In defining a 
schema we {\it average} over all strings that contain the given 
schema. In such a sum we are summing over all possible values for 
the string bits ${\C_i-\xi}$ present in the population. A schema
thus represents a coarse grained degree of freedom because we are
forfeiting explicit information about the out of schema string 
bits. Clearly the lower the order of the schema the higher the 
degree of coarse graining, the maximal coarse graining being 
associated with the maximally degenerate schema where $N_2=1$. 

A schema of order $N_2$ 
has only $N_2$ degrees of freedom and $2^{N_2}$ possible states.
Given that one of the fundamental characteristics of complex systems 
is the existence of a large number of degrees of freedom and an 
exponentially large state space any methodology that purports to 
reduce the number of EDOF will prove very 
useful. To see this in the context of an explicit example let us say 
that we wish to calculate the average fitness in a GA evolving according
to proportional selection with strings 
of size $N$, where $N$ is a multiple of $2$. The evolution equation
for the expected number of strings of type $\C_i$, $n(\C_i,t)$, is
\be
n(\C_i,t+1)={f(\C_i,t)\o\fav}n(\C_i,t).\label{nevol}
\ee
The average population fitness, $\fav$, for the case of a non-time dependent
landscape obeys the equation
\be
{\bar f}(t+1)=\sum\limits_{\C_i}{f^2(\C_i)\o\fav}P(\C_i,t)\label{fevol}
\ee
where $P(\C_i,t)=n(\C_i,t)/n$, $n$ being the population size which
we regard as being constant. As proportional selection is a 
stochastic process, for small population sizes one will typically
see large fluctuations, i.e. in any given experiment one may well
see large deviations between the results of (\ref{nevol}) and (\ref{fevol})
and the corresponding experimental quantities. However, taking averages
over repeated experiments the results converge to those of the above
equations. In fact, in the infinite population limit $P(\C_i,t)$ will converge
to the probability of finding string $\C_i$ at time $t$.
The string fitness maps 
every string state to $R^+$. If the population is large then many strings 
will be represented and hence many terms in $\fav$ will 
be non-zero. Thus, to calculate the evolution of $\fav$ one needs to
solve $\sim 2^N$ coupled equations. Let us instead take the 
following approach: we will average over odd string positions in the
population leaving strings, $\C'_i$, (or rather now schemata) of $N/2$ 
definite bits that satisfy
\be
n(\C'_i,t+1)={f(\C'_i,t)\o\fav}n(\C'_i,t)
\ee
where now the fitness of $\C'_i$ depends on time even if $f(\C_i)$ 
didn't. To calculate $\fav$ one now only needs to solve $2^{N/2}$ 
coupled equations. One can repeat this process, each step of coarse 
graining reducing the number of EDOF by one half, until we reach
the situation where a $1$-schema has been reached. This cannot be 
further coarse grained of course, except by passing to the trivial 
situation wherein all string bits are summed over, i.e. a 
$0$-schema. At this level the evolution equation for the
effective string of size one ($1$-schema $\alpha$) is
\be
n(\alpha,t+1)={f(\alpha,t)\o\fav}n(\alpha,t)
\ee
Now, $\alpha$ can only take two values, $1$ and $0$ say, hence
$\fav=[f(1,t)n(1,t)+f(0,t)n(0,t)]/n$. Thus, the problem of finding 
the average fitness has been reduced to that of solving a problem with one 
degree of freedom and two possible states!

So what's the catch? The principal, and more
fundamental, problem is that the genetic operators, principally reproduction and
crossover, are defined at the microscopic level. In other words, as can be 
seen in (\ref{scfit}), to assign a fitness to a schema one has to sum over
the different strings in the population that contain it. Thus, to calculate
quantities associated with the coarse grained degrees of freedom one must 
consider the microscopic degrees of freedom. One might be tempted therefore to 
think that even though there is an apparent reduction in the number of EDOF 
the net gain is canceled out by the fact that one has to return
to the microscopic degrees of freedom in order to calculate their 
evolution under the genetic operators. If one wished to calculate 
the dynamics exactly then the above would be true. However, returning 
to the idea of emphasizing the relevant degrees of freedom 
it may well be that in the averaging process 
certain ones are more important than others, therefore allowing one
to neglect, or treat as a perturbation, the effect of the irrelevant 
ones. In particular, near a fixed point of the dynamics one might 
well expect to see a simplification.

A second problem is that if we wish to ask a question
about a particular string and we only have access to schemata of 
order $N_2<N$ then the question will
be impossible to answer. In other words, if we are going to accept a coarse grained
description then we can only ask questions about coarse 
grained variables. This in no way will affect the calculation of 
population variables such as average population
fitness, standard deviation about the average fitness etc. 
Neither should it affect the ability of the GA to find an optimum as 
a fixed point of the dynamics as this can be represented in terms of 
optimal schemata.

In the above we discussed a particular coarse graining which led to 
a certain, definite set of schemata of order $N/2$, $N/4,...,1$ 
associated with averaging over the odd bits of each successive 
coarse grained string. Generally there are very many different 
coarse grainings possible, $3^N-1$ for a given string.
Which are useful and which aren't? This depends on the fitness landscape
under consideration. What one wishes to do is to choose a coarse
graining that gives rise to EDOF that are relatively weakly coupled.
Finding such a coarse graining may well of course be very difficult.
The coarse graining by factors of $2$ above is a proposal for 
an algorithm to calculate GA evolution. Whether this particular
coarse graining would be useful, as mentioned, depends on the
fitness landscape. Although the method might seem somewhat artificial it
is important to emphasize that such methods, based on the idea of 
the renormalization group (see for example Goldenfeld (1989) for 
a review) have proved to be extremely effective in many
areas of physics and applied mathematics and have yielded very good results
on canonical optimization problems such as the Traveling Salesman problem.

Later we will emphasize that $1$-schemata are very useful coarse grained 
variables, as being of size $1$ they are immune to the effects of 
crossover. In terms of $1$-schemata the average fitness in the population is
\be
\fav=\sum\limits_{i=\alpha}^N f(\alpha,t)P(\alpha,t)
\ee
where the sum is over the $N$ possible $1$-schemata, $f(\alpha,t)$ is 
the fitness of the $1$-schema $\alpha$ at time $t$ and 
$P(\alpha,t)=n(\alpha,t)/n$ is 
the expected proportion of strings present in the population that contain
the $1$-schemata $\alpha$. 
 
\section{ Schema Equation} 

In this section we will review the derivation of the schema 
evolution equation of 
\cite{stewael}, \cite{stewaell}. 
Given that the microscopic degrees of freedom are strings however, 
we will first derive an equation for strings evolving under  
the effects of the three genetic operators: proportional selection,
crossover and mutation. We will throughout only consider simple
one-point crossover. The analysis can be repeated, with analogous 
results, for the case of $n$-point crossover. 

We will consider the change in the expected number, $\nx$, 
of strings that contain a particular schema $\xi$, of order $N_2$ 
and length $l\geq N_2$, as a function of time (generation). 
If mutation is carried out after crossover one finds 
that the expected relative proportion of $\C_i$ in the population, 
$\pmic=n(\C_i,t)/n$, satisfies 
\be
\pmicl=\muti\pmicc + {\ds\sum_{\ss C_j\neq C_i}}\mutji\pmicjc
\label{mastcrossmut}
\ee
where the effective mutation coefficients are:
$\muti=\prod_{\s k=1}^{\ss N}(1-p_m(k))$, which is the probability that 
string $i$ remains unmutated, and $\mutji$, the probability that string $j$ 
is mutated into string $i$ given by 
\be
\mutji=\prod_{\ss{k\in \{C_j-C_i\}}}{p_m(k)}\prod_{\ss{k\in\{C_j-C_i\}_c}}
{(1-p_m(k))}
\ee
where $p_m(k)$ is the mutation probability of 
bit $k$. For simplicity we assume it to be constant, though the equations are 
essentially unchanged if we also include a dependence on time. 
$\s\{C_j-C_i\}$ is the set of bits that differ between $\C_j$ and 
$\C_i$ and $\s\{C_j-C_i\}_c$, the complement of this set, is the set 
of bits that are the same. In the limit where $p_m$ is uniform, 
$\muti=(1-p_m)^{\s N}$ and 
$\mutji=p_m^{\s d^{\ss H}(i,j)}(1-p_m)^{\s N-d^{\ss H}(i,j)}$,
where $\hamij$ is the Hamming distance between the strings $\C_i$ and $\C_j$.
The quantity $\pmicc$ is the expected proportion of strings of type $\C_i$ in
the population after selection and crossover. Explicitly
\be
\pmicc = \pmicf
-{p_c\o N-1}{\ds\sum_{\ss C_j\neq C_i}}{\ds\sum_{k=1}^{\ss N-1}}
\crossij\pmicf\pmicjf \\ 
+{p_c\o N-1}\sum_{\ss C_j\neq C_i}\sum_{\ss C_l\neq C_i}\sum_{k=1}^{\ss N-1}
\crossjl\pmicjf P'(\C_l, t)\nn
\ee
where $p_c$ is the crossover probability, $k$ is the crossover point, 
and the coefficients $\crossij$ and $\crossjl$,  
represent the probabilities that, given that $\C_i$ 
was one of the parents, it is destroyed by the crossover process, and
the probability that given that neither parent was $\C_i$ it 
is created by the crossover process. Explicitly
\be
\crossij=\theta(\hamijL)\theta(\hamijR)\label{thetas}
\ee
and
\be
\crossjl={1\o2}[\delta(\hamijL)\delta(\hamilR)
+\delta(\hamijR)\delta(\hamilL)]\label{deltas}
\ee
where $\hamijR$ is the Hamming distance between the right 
halves of the strings $\C_i$ and
$\C_j$,  ``right'' being defined relative to the crossover point, with 
the other quantities in (\ref{thetas}) and (\ref{deltas}) being similarly defined. 
$\theta(x)=1$ for $x>0$ and
is $0$ for $x=0$, whilst $\delta(x)=0\ \ \forall x\neq0$ and $\delta(0)=1$. 
Finally, $\pmicf=(\fmic/\fav) P(\C_i,t)$ gives the expected proportion of
strings $\C_i$ after the selection step. Note that $\crossij$ and 
$\crossjl$ are properties of the crossover process itself and therefore 
population independent. The equation 
(\ref{mastcrossmut}) yields an exact expression for the expectation 
values, $n(\C_i,t)$, and in the limit $n\ra\i$ yields the correct 
probability distribution governing the GA evolution. 

The evolution equation we have derived takes into account exactly the 
effects of destruction and reconstruction of strings and, at least 
at the formal level, has the same content as other exact formulations 
of GA dynamics 
\cite{vose}. 
It should also be formally equivalent to the equation of Bridges and Goldberg 
\cite{goldbridge}. 
Before passing
to the case of schemata it is interesting to put the equation into a simpler
form. To see this, consider first the destruction term. The matrix (\ref{thetas}) 
restricts the sum to those $\C_j$ that differ from $\C_i$ in at least one bit both
to the left and to the right of the crossover point. One can convert the sum over
$\C_j$ into an unrestricted sum by subtracting off those $\C_j$ that have 
$d^H_L(i,j)=0$ and/or $d^H_R(i,j)=0$. Similarly one may write the 
reconstruction term as
\be
\sum_{\ss C_j\supset C_i^L}\sum_{\ss C_l\supset C_i^R}
\pmicjf P'(\C_l, t)
\ee
where $\C_i^L$ is the part of $\C_i$ to the left of the crossover point and
correspondingly for $\C_i^R$. However, by definition
\begin{eqnarray}
{\bar f}(\C_i^{\ss L},t)= {\sum_{\ss C_j\supset C_i^L}f(\C_j,t)n(\C_j,t)  \o
\sum_{\ss C_j\supset C_i^L}n(\C_j,t)}
\end{eqnarray}
where $\sum_{\ss C_j\supset C_i^L}n(\C_j,t)$ is the total 
number of strings in the population that contain $\C_i^L$. The final 
form of the string equation without mutation thus becomes
\be
\pmicl=\pmicf
-{p_c\o N-1}\sum_{k=1}^{\s N-1}(\pmicf-\pmicLf\pmicRf)\label{stringfin}
\ee
with 
\be\pmicLf=\sum_{\ss C_j\supset C_i^L}\pmicjf\ee
and similarly for $\pmicRf$. It is important to note here that in this
form the evolution equation shows that crossover explicitly introduces
the idea of a schema and the consequent notion of a coarse graining.
$\C_i^L$ and $\C_i^R$ are schemata of order and length $k$ and $N-k$ 
respectively. 

The analogous equation for schema evolution can be found by summing
equation (\ref{mastcrossmut}) over all strings that contain the schema of 
interest $\xi$. The result is
\be
\pxl=\mutis\pxc + \sum_{\s \slashxi_i}\mutijs\pnxc\label{maseqtwo}
\ee
where
\be
\pxc= \pxf - {p_c\o N-1}\sum_{k=1}^{l-1}\left(\pxf-\pxLf\pxRf\right)
\ee
and the sum in (\ref{maseqtwo}) is over all schemata $\slashxi_i$ that differ
by at least one bit from $\xi$ in one of the $N_2$ defining bits of $\xi$.
All other quantities are the schema analogs of quantities defined in 
(\ref{mastcrossmut}). The effective mutation coefficients 
$\mutis$ and $\mutijs$ are 
\be
\mutis=(1-p_m)^{N_2} \ \ \ \ \ \
{\rm and} \ \ \ \ \ \ 
\mutijs=p_m^{d^H(\xi,\slashxi_i)}(1-p_m)^{N_2-d^H(\xi,\slashxi_i)}
\ee
where $d^H(\xi,\slashxi_i)$ is the Hamming distance between the schemata
$\xi$ and $\slashxi_i$.

A very interesting feature of the evolution equations we have 
presented is their form invariance under a coarse graining. Starting 
with the string equation any coarse graining to schemata of order 
$N_2<N$ yields an equation identical in form to that of its 
predecessor. 

\section{Effective Fitness}

Having derived the schema evolution equation, before turning
to an analysis of its many features, we wish to digress on the notion
of fitness. The main intuitive idea behind fitness is that fitter parents
have more offspring. In equation (\ref{maseqtwo}), neglecting for the
moment mutation and crossover, taking the limit of a continuous
time evolution one finds
\be
P(\xi,t)=P(\xi,0){\rm e}^{\int_0^t s_{\ss \xi}dt'}
\ee
where $s_{\ss\xi}=\frat-1$ is the selective advantage of the schema $\xi$. 
If $s_{\ss\xi}>0$ the expected number of $\xi$ grows, whilst if 
$s_{\ss\xi}<0$ it decreases. However, consider the following two simple cases. 
First, consider the effect of mutation without crossover in the context of
a model that consists of $2$-schemata, $11$, $01$, $10$, $00$, where
each schema can mutate to the two adjacent ones when the states
$11$, $10$, $00$, $01$ are placed clockwise on a circle. 
For example, $11$ can mutate
to $10$ or $01$ but not to $00$. This is evidently the limit where two-bit
mutations are completely negligible compared to one-bit mutations. 
We assume a simple degenerate fitness landscape: $f(11)=f(01)=f(10)=2$, $f(00)=1$.
In a random population, $P(11) = ... = P(00) = {1 \over 4}$. 
If there is uniform probability $p_m$ for each schema to mutate to an adjacent
one then the evolution equation that describes this system is
\be
P(i,t+1)=(1-2p_m)P'(i,t)+p_m(P'({i-1},t)+P'({i+1},t))
\ee
For $p_m=0$ the steady state population is $P(11)=P(01)=P(10)=1/3$, 
$P(00)=0$. Thus we see the synonym symmetry of the 
landscape associated with the degeneracy of the states $11$, $10$ and $01$
is unbroken. However, for $p_m>0$, the schemata distribution at $t=1$, 
starting from a random distribution at $t=0$, is $P(11)=2/7$, 
$P(01)=P(10)=(2-p_m)/7$, $P(00)=(1+2p_m)/7$.
Thus, we see that there is an induced breaking of the landscape synonym 
symmetry due to the effects of mutation. In other words the population 
is induced to flow along what in the fitness landscape is a flat direction. 

As a second example consider the $2$-schemata problem now with crossover 
but neglecting mutation, and with a fitness landscape where $f(01)=f(10)=0$ and
$f(11)=f(00)=1$. The steady state solution of the schema evolution equation is
\be
P(11)=P(00)={1\over2}\left(1-{p_c\over2}{(l-1)\over(N-1)}\right)\quad\quad
P(01)=P(10)={p_c\over4}{(l-1)\over(N-1)}
\ee
For $l=N$ and $p_c=1$ we see that half the steady state population is composed
of strings that have zero fitness!

Although the above examples are artificial they serve to make the point that
the genetic operators can radically change the ``effective'' landscape in
which the population evolves. The actual ``bare'' landscape associated
purely with selection in the above offers very little intuition as to the
true population evolution. Real populations can flow rapidly along flat
directions and strings may be present even if they have zero fitness. To take
this into account we propose using an ``effective'' fitness function
\cite{stewael}, \cite{stewaell} 
defined via
\be
\pxl={\ef\o\fav}\px\label{three}
\ee
comparing with equation (\ref{maseqtwo}) one finds
\begin{eqnarray}
\ef=\mutis\fxi+\sum_{\s \slashxi_i}\mutijs{\pnx\o\px}{\bar f}(\slashxi_i,t)\nn\\
-{p_c\o N-1}\mutis\fav \sum_{k=1}^{N-1} \left({P'(\xi, t)
-P'(\xi_{L}, t)P'(\xi_{R}, t) \o \px} \right)\nn\\
-{p_{\ss c} \o N-1}\sum_{\s \slashxi_i}\mutijs
\fav \sum_{k=1}^{N-1} \left({P'(\slashxi_i, t)
-P'(\slashxi_{i_L}, t)P'(\slashxi_{i_R}, t) \o \px} 
\right)
\label{effit}   
\end{eqnarray}
In the limit $p_m\ra0$, $p_c\ra0$ we see that $\ef\ra\fxi$.
The above also leads to the idea of an effective selection 
coefficient, $s_{\ss\rm eff}=\ef/\fav-1$, that measures directly selective
pressure. If we think of $s_{\ss\rm eff}$ 
as being approximately constant in the vicinity of time $t_{\ss 0}$, 
then $s_{\ss\rm eff}(t_0)$ gives us the exponential rate of increase or 
decrease of growth of the schema $\xi$ at time $t_0$.
In the limit of a continuous time evolution the solution of 
(\ref{three}) is
\be
P(\xi,t)=P(\xi,0){\rm e}^{\int_0^ts_{\ss\rm eff}dt'}\label{four}
\ee

In the case of the toy examples above: for mutations without crossover
\be
f_{\ss {\rm eff}}(i,t) = f_i+{p_m\over P(i,t)}(f_{i-1}P({i-1},t)+f_{i+1}P({i+1},t)-2f_iP(i,t))
\ee
At $t=0$, $f_{\ss {\rm eff}}(11,0)=2$, $f_{\ss {\rm eff}}(01,0)=f_{\ss {\rm eff}}(10,0)=2-p_m$ and
$f_{\ss{\rm eff}}(00,0)=1+2p_m$. Thus we see that the 
effective fitness function provides a selective pressure by selecting among the 
degenerate schemata those that have a higher probability 
to produce fit descendents. 

Of course, the definition of effective fitness is not unique. Another
natural definition follows from the split into those terms of the
evolution equation that are linear
in $P(\xi,t)$ and those ``source'' terms that are independent of it.
For instance, in the case of selection and crossover we have 
\be
\pxl= {\efff\o\fav}\px+j(t)\label{effeq}
\ee
where $\efff = (1- p_c{(l-1)\o N-1}){\fxi\o\fav}$ and 
$j(t)={p_c\o N-1}\sum_{k=1}^{l-1}\pxLf\pxRf$. The corresponding
effective selection coefficient is 
$s'_{\ss{\rm eff}} = ((1- p_c{(l-1)\o N-1}){\fxi\o\fav}-1)$. In the limit
of a continuous time evolution (\ref{effeq}) may be formally integrated to yield
\be
P(\xi,t)={\rm e}^{\int_0^t s'_{\ss{\rm eff}}(t')dt'}P(\xi,0)+
{\rm e}^{\int_0^t s'_{\ss{\rm eff}}(t')dt'}
\int^t_0j(t'){\rm e}^{-\int_0^{t'} s'_{\ss{\rm eff}}(t'')dt''}dt'
\label{formsol}
\ee

\section{Schema Theorem and Building Blocks}

We now turn to a discussion of the schema theorem and the building 
block hypothesis. The standard ``schema theorem'' 
\cite{holland,goldberg}, 
or fundamental theorem of GAs, states that for a schema, $\xi$, of length $l$
evolving according to proportional selection and $1$-point crossover
\be
\pxl\geq\pxf\left(1-p_c\left({l-1\o N-1}\right)-N_2p\right),\label{schth}
\ee
and has the interpretation that schemata of higher than average 
fitness will be allocated exponentially more trials over time. 
The conventional schema theorem only provides us with a lower bound 
for the expected number of schemata due to the fact that it does not
explicitly account for schema reconstruction. 
Equation (\ref{maseqtwo}), however, exactly takes into account the 
effect of schema reconstruction due to both mutation and 
crossover. Together with the definition of effective fitness 
in equation (\ref{effit}) of the previous section it allows one to 
state a new schema theorem:
\vskip 0.3cm

\noindent{\bf Schema Theorem}
\be
\pxl={\ef\o\fav}\px\label{schtheorem}
\ee
\vskip 0.3cm
The interpretation of this equation is clear and analogous to the 
old schema theorem: schemata of higher 
than average {\it effective fitness} will be allocated an ``exponentially'' 
increasing number of trials over time. We put the word exponentially
in quotes as the real exponent, $\int^t s_{\ss{\rm eff}}dt'$, is not, 
except for very simple cases such as a flat fitness landscape,
of the form $\alpha t$, where $\alpha$ is a constant. The 
illustrative examples of the last section show that there is 
potentially a strong difference between the standard selection based
fitness and effective fitness as the latter takes into account the 
effect of all genetic operators. The fact that strings with zero 
selective fitness can receive an exponentially increasing number of 
trials shows quite clearly that effective fitness is a more relevant 
concept. In this sense our schema theorem does not just state the 
obvious --- that fit schemata that are preserved by the crossover
operator will prosper. Once again this emphasizes the role of the 
destructive effect of crossover. The novel element here is seeing exactly
how important is schema reconstruction. In fact generically it is the 
dominant contribution.  

The schema evolution equation we have derived possesses many interesting
features one of the most interesting being the way that it relates
evolution in time to different levels of coarse graining. To see this
we first return to the string evolution equation (\ref{mastcrossmut}).
Up to this point we have presented our results in almost the most
general way possible --- for any type of landscape and taking into account
both crossover and mutation. Throughout the rest of the paper we 
will concentrate more on the effect
of crossover and thus neglect mutation. The reason for this is that
we will mainly be concerned with the importance of schema length vis a vis
the building block hypothesis. Mutation being a strictly local operator
will not play a major role in this discussion.
Note that this equation is written entirely in terms of the fundamental
degrees of freedom --- the strings. In passing to the form (\ref{stringfin})
we have performed a coarse graining by summing over all strings that
contain $\C_i^L$ irrespective of what lies to the right of the crossover
point, and similarly for strings containing $\C_i^R$. The implication
is that the very nature of crossover imposes on us the idea of coarse
graining, and more specifically the idea of a schema, given that 
$\C_i^L$ and $\C_i^R$ define schemata of order and size $k$ and $N-k$
respectively. In order to solve the equation (\ref{stringfin}) we
need to know $\pmicL$ and $\pmicR$. However, these in turn obey
evolution equations of the form
\be
P({\C_i^L},t+1)=\pmicLf
-{p_c\o N-1}\sum_{m=1}^{\s k-1}(\pmicLf-\pmicLLf\pmicLRf)
\ee
where $\C_i^{LL}$ and $\C_i^{LR}$ are the left and right parts of 
$\C_i^L$, left and right being defined relative to the crossover point $m$,
where $m<k$. Now, $\C_i^{LL}$ and $\C_i^{LR}$ as schemata
are more coarse grained than $\C_i^L$, i.e. they are of lower order. 
Clearly this pattern of behavior continues, i.e. in order to
calculate $P(\C_i,t+1)$ one requires $P(\C_i^L,t)$ and $P(\C_i^R,t)$ which
in their turn require $P(\C_i^{LL},t-1)$, $P(\C_i^{LR},t-1)$, 
$P(\C_i^{RL},t-1)$ and $P(\C_i^{RR},t-1)$ etc. 
For each step back in time we pass to 
more coarse grained degrees of freedom. ${\C_i}$ thought of as a schema
is of higher order than $\C_i^L$ or $\C_i^R$, 
which in their turn are of higher
order than $\C_i^{LL}$, $\C_i^{LR}$, $\C_i^{RL}$ 
and $\C_i^{RR}$. So where 
does this process stop? The maximally coarse grained EDOF are $1$-schemata.
It is not possible to cut a $1$-schemata and hence crossover is explicitly 
neutral, i.e. $1$-schemata obey the equation
\be
P(i,t+1)=P'(i,t)
\ee
As a simple example consider a $4$-bit string $ijkl$. The hierarchical structure
of one possible ancestral tree can be written as
$$
\displaylines{
\rlap{t+1} \hfill ijkl \hfill\cr
\rlap{t} \hfill ijk,l \quad\quad ij,kl \quad\quad i,jkl \hfill\cr
\rlap{t-1} \hfill ij,k \quad i,jk \quad\quad i,j \quad k,l \quad\quad jk,l \quad j,kl \hfill\cr
\rlap{t-2} \hfill i,j \quad\quad j,k \quad\quad\quad\quad\quad\quad j,k \quad\quad k,l \hfill\cr}
$$ 
This tree shows only the effect of the reconstruction term in the schema 
equation over the space of 3 generations. Of course there are many other 
processes that contribute to the appearance of $ijkl$ at time $t+1$ that involve
various combinations of schemata destruction and reconstruction. As far
as pure schemata reconstruction is concerned however we see 
that $1$-schemata play a privileged role as they represent the ultimate 
building blocks. For an $N$-bit string the maximum number
of time steps before all ancestors are $1$-schemata is $N-1$.

All the above equally applies to a generic schema, $\xi$, composed of 
schemata, $\xi_L$ and $\xi_R$ which in their turn are composed of 
the schemata $\xi_{LL}$, $\xi_{RL}$, $\xi_{LR}$ and $\xi_{RR}$ etc. It
should be clear that the idea of building blocks is manifest in the very
structure of our evolution equations. $\xi_{LL}$, $\xi_{RL}$, $\xi_{LR}$ and 
$\xi_{RR}$ are building blocks for $\xi_L$ and $\xi_R$ which in their
turn are building blocks for $\xi$. The ultimate building blocks are of course
the $1$-schemata. In the above example of a $4$-bit string or schema the
four building blocks of order one, $i$, $j$, $k$ and $l$ combine to form
building blocks of order two $ij$ and $kl$ which in turn combine with the
building blocks of order one to form building blocks of order three, $ijk$ and
$jkl$. The building blocks of order three combine with the blocks of order one
and the blocks of order two combine together to give blocks of order four, and
so the process continues.

In terms of the effective fitness, $\efff$ introduced previously
\be
P(\xi,t)={\rm e}^{\int_0^t \seff'(t')dt'}P(\xi,0)+
 {p_c{\rm e}^{\int_0^t \seff'(t')dt'}\o N-1}\sum_{k=1}^{N-1}\int^t_0\pxLft\pxRft
{\rm e}^{-\int_0^{t'} \seff'(t'')dt''}dt'
\label{formsol2}
\ee
Up to now we have been able to analyze a general landscape. To arrive at 
more explicit, analytic formulae in an arbitrary landscape
is prohibitively difficult.  We will therefore temporarily restrict
our attention to some more restrictive but simpler cases. 
We start with the case of a flat fitness landscape. In this
case $\seff=-p_c(l-1)/(N-1)$, hence
\be
P(\xi,t)={\rm e}^{-p_c{(l-1)\o(N-1}t}P(\xi,0)+
{\rm e}^{-p_c{(l-1)\o(N-1)}t}{p_c\o N-1}\sum_{k=1}^{N-1}\int^t_0\pxLft\pxRft
{\rm e}^{p_c{(l-1)\o(N-1)}t'}dt'
\ee
Notice that dependence on the initial condition, $P(\xi,0)$, is exponentially
damped unless $\xi$ happens to be a $1$-schema, the solution of the 
$1$-schemata equation being
\be
P(i,t)=P(i,0)
\ee
An immediate consequence is that when considering the source term describing
reconstruction the only non-zero terms that need to be taken into account are
those which arise from $1$-schemata, as any higher order term will always
have an accompanying exponential damping factor. Thus we see that the fixed
point distribution for a GA with crossover evolving in a flat fitness landscape
is
\be
P^*(\xi)=Lt_{t\ra\i}P(\xi,t)=\prod_{i=1}^{N_2}P(i,0)
\ee
which is basically Geiringer's Theorem 
\cite{geir} 
in the context of schema
distributions and simple crossover. We see here that the theorem appears in 
an extremely simple way as a consequence of the solution of the evolution equation.

Note that this fixed point distribution arises purely 
from the effects of reconstruction,
the absence of which leads to a pure exponential damping and the unphysical
behavior $P(\xi)\ra0$. We can also see from the above that a version of 
Geiringer's theorem will also hold in a more general non-flat landscape 
where selection is 
only very weak, where what we mean by weak is that $\frat\sim(1+\epsilon)$ and 
$\epsilon<{p_c{(l-1)\o (N-1)}\o 1-p_c{(l-1)\o (N-1)}}\ \  \forall l>1$. 
Under such circumstances
once again anything other than a $1$-schema will be associated with an
exponential damping factor. A distinction between the two cases however
is that for a flat fitness landscape the fixed point is fixed by the initial 
proportions of the various $1$-schemata as there is no competition between them.
Here, however, due to the non-trivial landscape certain 
$1$-schemata are preferred 
over others. A concrete example of such a landscape would be
$f_i=1+\alpha_i$ where $\sum_i\vert\alpha_i\vert\leq{\epsilon\o(2+\epsilon)}$ 
and $f_i$ is the fitness of the ith bit. 
Note there is no need to restrict to a linear fitness function here, arbitrary epistasis
is allowed as long as it does not lead to large fitness deviations away from the mean.
In this case $\frat<1+\epsilon$. 

So what is the analog of the building block hypothesis here? Our schema theorem 
states that schemata of above average {\it effective} fitness will be allocated
``exponentially'' more trials over time. In the way the evolution 
equation is structured we see that the effective fitness in terms of 
the effects of crossover consists of a
destruction term and a reconstruction term. Inherent in the structure 
of the reconstruction term is a form of the building block 
hypothesis --- that higher
order schemata are built from fit, shorter, lower order schemata. If 
$\pxf>\pxLf\pxRf$ then the effects of destruction will outweigh those of
reconstruction, whilst if $\pxf<\pxLf\pxRf$ reconstruction will dominate. 
The content of this inequality is that reconstruction will dominate destruction
if the probability to select the parts of a schema is greater than the probability
to select the whole schema. Once again this is a general conclusion valid for
any landscape. To give a more analytic slant we restrict to 
the case of two-schemata in a flat fitness landscape wherein one finds
\be
s_{\ss{\rm eff}}=-p_c\left({l-1\o N-1}\right)+p_c\left({l-1\o N-1}\right){P(i,0)P(j,0)\o P(ij,t)}
\ee
Thus we see that the effect of reconstruction is greater than that of destruction
if $i$ and $j$ are negatively correlated. Notice that if reconstruction
is more important then the contribution from the latter is maximized by
maximizing the schema length, $l$. In other words large, rather than small, 
schemata are favored!

In general the fitness landscape itself induces correlations between 
$\xi_L$ and $\xi_R$. In this case there is a competition 
between the (anti-) correlating effect of the landscape and 
the mixing effect of crossover. Selection itself more often than not 
induces an {\it anti}-correlation between fit schemata parts, rather than
a positive correlation. Indeed, in the neutral case of a non-epistatic landscape 
one has $1 + {2N_2 \o N} \dfx < (1 + {2N_L \o N} \dfxL)
(1 + {2N_R \o N} \dfxR)$ where $\dfx$, $\dfxL$ and $\dfxR$ are the fitness
deviations of the schemata $\xi$, $\xi_L$  and $\xi_R$ from an average fitness
which we have normalized to one half. Thus we see that selection induces an 
anti-correlation when $\dfxL, \dfxR > 0$ and hence in an uncorrelated initial population, 
$P'(\xi, t) < P'(\xi_L, t) P'(\xi_R,t)$. This means that crossover 
plays an important role in allowing both parts of a successful schema to
appear in the same individual. The effect of crossover is to weaken but 
not cancel completely the anti-correlations induced by  selection
and thus make it easier to find the whole schema. 
Indeed, it is possible to show that for a non-epistatic landscape that
the contribution to population fitness from all schemata of length $l$,
starting with a random initial population at time $t$, is independent of $l$ at
time $t+1$ and is an increasing function of $l$ for large $l$ at 
time $t+2$ 
\cite{stewael}.

More complicated landscapes one has to examine on a case by case basis.
Of course, it is always possible to invent a landscape where there is a
fitness advantage associated with bits that are close together. However, it
is equally easy to find one where there is a fitness advantage for bits that
are widely separated. The non-epsitatic landscapes above 
are neutral in this respect
and therefore any results about the nature of schemata and building blocks
are a reflection of the geometric effect of crossover and not associated with
bit-bit correlations induced by the landscape itself. We now have ample 
experimental evidence that this is the case as well for ``generic'' landscapes
with epistasis such as the Kaufmann $Nk$ models. This evidence will be 
published elsewhere. A particularly interesting example of epistasis
is deception as it has played an important role in the 
theory of GAs 
\cite{gold87}. 
The very nature of deception is such that the
bits of a schema are less selected than the whole and hence we can see
from the schema equation that in this circumstance destruction 
will outweigh reconstruction. 
However, this will only be totally deceptive if all possible schema reconstruction 
channels $\xi_L+\xi_R\ra\xi$ are deceptive. For a schema of order $N_2$ there
are $N_2-1$ such channels. Thus, for $N_2$ large it will 
typically be quite unlikely that
all channels will be deceptive. If there exist non-deceptive channels then it 
is probable that the population will evolve in those directions. In fact, 
as the example of a two-schema shows, for every deceptive channel there
is a non-deceptive one. One may explicitly see this from
\be
P(11,t+1)=P'(11,t)-p_c\left({l-1\o N-1}\right)(P'(11,t)P'(00,t)-P'(01,t)P'(10,t))\nn\\
P(01,t+1)=P'(01,t)-p_c\left({l-1\o N-1}\right)(P'(01,t)P'(10,t)-P'(11,t)P'(00,t))
\ee
Here $11$-channel deception, i.e. $P'(11,t)P'(00,t)>P'(01,t)P'(10,t)$, implies 
that the $01$-channel is non-deceptive. However, this is not much consolation
if the $11$-schemata happens to be the optimum. 
If we start with a random population then $11$-channel deception is 
equivalent to the statement $f_{\ss{\rm eff}}(11)<f(11)$.
For something as 
simple as the two-schemata problem there is only a single $11$-channel. For 
the $4$-bit schemata $ijkl$ we see that there are six reconstruction channels
in total. There are various ways to end at a totally deceptive problem. For 
instance,  the three channels $ijk+l\ra ijkl$, $ij+kl\ra ijkl$ and $i+jkl\ra ijkl$ 
might all be deceptive. Alternatively all the $1$-schemata $\ra$ $2$-schemata
channels might be deceptive. Generically the deviation of the effective fitness 
from the selective fitness will offer a reasonable measure of deception.

\section{Conclusions}

In this paper we have analyzed the Schema Theorem and the 
Building Block Hypothesis based on an exact evolution
equation for GAs. At the level of the microscopic degrees
of freedom, the strings, we established that the action 
of crossover by its very nature
introduces the notion of a schema, the probability to 
reconstruct a given string being dependent on the probabilities
for finding the right and left parts of the string relative to the 
crossover point in the two parents. These probabilities involve a
coarse graining, i.e. an averaging over all strings that contain
the constituent parts of the string, and hence represent schema
probabilities. We saw that the same equation, after a suitable 
coarse graining, also described the evolution of any arbitrary
schema.

One might enquire as to what advantages a formulation based 
on schemata, as has been
presented here, has over other existing formulations, such as the 
Vose Markov chain model. Indeed, the value of schemata and the 
Schema theorem in understanding GA evolution has been seriously
questioned 
\cite{gref,vos,rad,muhl}
There are many possible answers to this question: first a pragmatic
one --- that all ``things''
are made out of ``building blocks'', whether they be tables, giraffes or
computer programmes. Having an exact, amenable description of complex 
systems from the microscopic point of view is a vain hope. Complex systems
and complex behaviour can much better be understood in terms of 
EDOF. EDOF, almost by definition, are much fewer in number than the 
microscopic degrees of freedom and hence, in principle, would offer
a computationally simpler picture. However, the number of ways of 
combining the microscopic degrees of freedom into EDOF is very large,
hence one might think that such a description is even more costly than
one based on the microscopic degrees of freedom such as the Vose model.
This would be true if in analysing the GA one had to search through
all the possible ``coarse grainings'' available. For a given landscape, however,
a preferred coarse graining will often suggest itself. Secondly, 
we believe strongly that approximation schemes for solving GA evolution
equations will be much more forthcoming via a formulation in terms of 
schemata wherein one may appeal to all the intuition and machinery
of the renormalization group.

We introduced the notion of effective fitness showing through explicit 
examples that it was a more relevant concept than pure selective fitness
in governing the reproductive success of a schema. Based on this concept
of effective fitness and our evolution equation we introduced a new 
schema theorem that showed that schemata of high effective fitness 
received an exponentially increasing number of trials as a function of time. 
We then went on to discuss the building block hypothesis. One of the more
remarkable features of our equation is that it implicitly contains a version
of the latter in that the structure of the reconstruction
term relates in an ancestral tree the relation between a given schema
and its more coarse grained ancestors as a function of time. This ancestral
tree terminates at $1$-schemata, which are in some sense the ultimate 
building blocks as they cannot be destroyed by crossover. We also showed that
generically there is no preference for short, low-order schemata. In fact if
schema reconstruction dominates the opposite is true, typically
large schemata will be favored. Only in deceptive problems does it generally
seem that short schemata will be favored, and then only in totally deceptive
problems as the system will tend to seek out the non-deceptive channels
if they exist.

There are many points of departure from the present work to future 
research. On the theoretical side it will be very interesting to see
if other exact results besides Geiringer's theorem follow very simply from 
our evolution equation. A fundamental issue is trying to find approximation
schemes within which the equations can be solved, as for a general landscape
an exact solution will be impossible. In this respect, as mentioned, techniques familiar
from statistical mechanics such as the renormalization group might well prove
very useful. In fact the very structure of our evolution equation is very
similar to that of a renormalization group equation, a theme we shall return to
in a future publication. It is of course necessary to verify the equations
numerically. Some work in this direction has already been done 
\cite{stewael}
and further work has confirmed its qualitative conclusions
\cite{stewaelag}. In this respect
one has to tread carefully, as the interplay between selection and crossover can
be very subtle as the work on Royal Road functions 
\cite{mitch} 
has shown. 
Although very simple we favor preliminary analytic analyses based on non-epistatic
landscapes where one knows that there is no intrinsic inter-bit linkage due to
the fitness landscape and therefore one can study the geometric effects of
crossover in a more uncluttered environment.

\subsubsection*{Acknowledgements} 

This work was partially supported through DGAPA-UNAM grant number IN105197.
CRS wishes to thank Drs. David Goldberg and Michael Vose for enlightening
discussions. We also thank the entire Complex Systems group at the ICN under NNCP
(http://luthien.nuclecu.unam.mx/nncp) for maintaining a stimulating 
research atmosphere.

\end{document}